\documentclass[runningheads,a4paper]{llncs}

\usepackage{amssymb}
\usepackage{graphicx}

\usepackage{url}
\urldef{\mailsc}\path|melezhik@theor.jinr.ru|
\newcommand{\keywords}[1]{\par\addvspace\baselineskip
\noindent\keywordname\enspace\ignorespaces#1}

\begin{document}

\mainmatter  

\title{Multi-Channel Computations in Low-Dimensional Few-Body Physics}

\titlerunning{Multi-Channel Computations in Low-Dimensional Few-Body Physics}

%
%
\author{Vladimir S. Melezhik}
\authorrunning{V.S. Melezhik}

\institute{Bogoliubov Laboratory of Theoretical Physics,\\
Joint Institute for Nuclear Research, 141980 Dubna, Russian Federation\\
\mailsc\\
\url{http://theor.jinr.ru/~melezhik/}}

%
%

\toctitle{Multi-Channel Computations in Low-Dimensional Few-Body Physics}
\tocauthor{V.S. Melezhik}
\maketitle

\begin{abstract}
In this lecture I give a brief review of low-dimensional few-body problems recently encountered in
attempting a quantitative description of ultracold atoms and molecules confined in 2D and 1D optical lattices.
Multi-channel nature of these processes has required the development of
special computational methods and algorithms which I discuss here as
well as the most interesting results obtained with the offered
computational technique and future perspectives.
\end{abstract}

\section{Introduction}

Recent advances in physics of ultracold quantum gases have opened a
unique opportunity to study of low-dimensional quantum systems (see, for example \cite{pitaevskii,chin}).
In experiments with ultracold atoms and molecules the particle motion is
``frozen" along one or two variables by using the optical potential and
its dynamics is limited to a plane or a line. In this case, the physical properties of the low-dimensional
system can be dramatically altered under the action of the optical
trap. Moreover, one can effectively manage interatomic interactions by varying the parameters of the trap.
This opens up new possibilities for studying strongly correlated quantum systems \cite{pitaevskii},
few-body and many-body effects in low-dimensional systems \cite{chin,bloch}, new mechanisms of molecule
formation \cite{kohler}, and possible elements for quantum computer \cite{pitaevskii}.

The simplest example of few-body systems, a quantum two-body system, acquires in a confined geometry
of a 1D trap unique properties as compare to the conventional two-body problem in a 3D free
space \cite{pitaevskii,chin}. Confining of the motion in the transversal directions leads
to transverse quantization of the motion and makes the pair collisions in
a 1D trap essentially multichannel ones. Possible transverse excitation energies of the particle define
the channel thresholds for the scattering in the trap. Depending on the initial conditions, the parameters of the
trap and the interatomic interaction, different multichannel effects may occur at pair atomic
collisions in the 1D confining trap. It is a confinement-induced resonance (CIR) defined for zero-energy limit when
the ground state energy threshold of the colliding pair coincides with the resonance
in the closed transverse channel \cite{olshanii,bergeman}, transverse
excitations/deexcitations during the collision \cite{saeidian2008}, center-of-mass
effects \cite{kim2006,melezhik2007,kestner}, a new mechanism for molecule formation \cite{melezhik2009},
a CIR-splitting in anisotropic traps \cite{haller},
anharmonicity effects \cite{kestner}, and three-body effects \cite{chin}.

A quantitative description of the mentioned effects has
demanded to develop new computational methods that we consider in
the next section. Some results obtained with the elaborated computational
techniques are briefly discussed in the section 3. Concluding remarks are
given in the last section 4.

\section{Methods}

\subsection{Quantum Two-Body Problem in Confined Geometry of 1D Harmonic Trap}

The collisional dynamics of two distinguishable atoms
with coordinates $\bf{r}_1,\bf{r}_2$ and masses $m_1, m_2$ moving in the harmonic waveguide with
the transverse potential $\frac{1}{2} \sum_i m_i \omega_i^2 \rho_i^2$ ($\rho_i = r_i sin \theta_i$)
is described by the 4D time-dependent Schr\"odinger equation ($\hbar =1$)
\cite{melezhik2007,kim2006}
$$
i\frac{\partial}{\partial t}\psi(\rho_R,{\bf r},t)=H(\rho_R,{\bf r})\psi(\rho_R,{\bf r},t)
$$
with the Hamiltonian
\begin{equation}
H(\rho_R,{\bf r})= H_{CM}(\rho_R) + H_{rel} ({\bf r}) +
W(\rho_R,{\bf r}) \,\,.
\end{equation}
Here
\begin{equation}
H_{CM}= -\frac{1}{2M}(\frac{\partial
^2}{\partial \rho_{R}^{2}} +\frac{1}{\rho_{R}^{2}}
\frac{\partial^2}{\partial \phi^2} +\frac{1}{4\rho_{R}^{2}})
+\frac{1}{2}(m_{1}\omega_{1}^{2}+m_{2}\omega_{2}^{2})\rho_{R}^{2}
\end{equation}
and
\begin{equation}
H_{rel} = -\frac{1}{2\mu}\frac{\partial ^{2}}{\partial r^{2}}+
\frac{L^{2}(\theta,\phi)}{2\mu
r^{2}}+\frac{\mu^2}{2}(\frac{\omega_{1}^{2}}{m_{1}}+\frac{\omega_{2}^{2}}{m_{2}})
\rho^2+V(r)
\end{equation}
describe the center-of-mass (CM) and relative atomic motions. The potential  $V(r)$ describes the atom-atom interaction,
$\rho_R$ and ${\bf r} = {\bf r_1} - {\bf r_2}\mapsto (r, \theta, \phi) \mapsto (\rho, \phi, z)$ are the polar
radial CM and the relative coordinates and
$M=m_1+m_2$, $\mu = m_1 m_2/ M$. The term $\frac{L^2(\theta,\phi)}{2\mu r^2}=
-\frac{1}{2\mu r^2\sin \theta}(\frac{\partial}{\partial\theta}\sin \theta \frac{\partial}{\partial\theta} +\frac{1}{\theta}
\frac{\partial^2}{\partial\phi^2})$
represents the angular part of the kinetic energy operator of the relative
atomic motion.

The term
\begin{equation}
W(\rho_R,{\bf r})=\mu(\omega_{1}^2-\omega_{2}^{2}) r \rho_{R} \sin \theta \cos \phi
\end{equation}
 in the Hamiltonian (1) leads for two distinguishable atoms that feel different confining
frequencies $\omega_1\neq\omega_2$ to a coupling of the CM and relative atomic
motion, i.e. to the nonseparability of the quantum two-body problem in confined geometry of the 1D
harmonic trap. The problem is to integrate the Schr\"odinger equation from time $t=0$ to the
asymptotic region $t\rightarrow +\infty$ with the initial wave-packet

\begin{equation}
\psi(\rho_{R},{\bf r},t=0)=N r \sqrt{\rho_R}
\exp\{-\frac{\rho_1^2}{2a_1^2} - \frac{\rho_2^2}{2a_2^2}
-\frac{(z-z_{0})^{2}}{2a_{z}^{2}}+ik_{0}z\}
\end{equation}
representing two different noninteracting atoms in the transversal ground
state of the waveguide with $a_i=(1/m_i\omega_i)^{1/2}$ and the overall
normalization constant $N$ defined by $< \psi(0)|\psi(0)> = 1$. We choose
$z_0\rightarrow -\infty $ to be far from the origin $z=0$ and
$a_z\rightarrow\infty$ to obtain a narrow width in momentum and energy
space for the initial wave-packet. The wave-packet moves with a positive
interatomic velocity $v_{0}=k_0/\mu=\sqrt{2\epsilon_{\|}/\mu}$ (defined by the longitudinal
colliding energy $\epsilon_{\|}$) to the scattering region at $z=0$ and splits up after the scattering into two parts
moving in opposite directions $z\rightarrow \pm\infty$. If the initial conditions permit opening inelastic channels it can lead to the
collisional excitation of the transverse vibrations\cite{saeidian2008} or formation of molecular bound states
\cite{melezhik2009}. Multichannel character of the scattering is also
developing at the region of the CIR where resonance in the closed channel
leads to the resonant behaviour of the elastic scattering amplitude if
the initial colliding energy coincides with the resonant energy of molecular state in the
closed channel \cite{bergeman,melezhik2007,saeidian2008}.

The modeling of the interatomic interaction $V(r)$ was discussed in \cite{bergeman,melezhik2007,melezhik2009}.

\subsection{Discretization of the Angular-Subspace: 2D Nondirect Product Discrete Variable Representation}
For discretization of the angular part in (1) we apply the 2D nondirect discrete variable
representation \cite{melezhik1997,melezhik1998,melezhik1999}.
For 1D quantum problems discrete variable representation (DVR) was suggested in \cite{light} and broadly applied
in the quantum chemistry computations \cite{carrington2000}. However, it happened that the attempts to extend this approach for
the case of more higher dimensions as a direct product of 1D DVRs were not so efficient \cite{carrington2000,carrington}.
In 1997 we have suggested a 2D nondirect product DVR (NDDVR) in application to the time-dependent 3D Schr\"odinger equation
describing hydrogen atom in a laser field of arbitrary polarization \cite{melezhik1997,melezhik1998}. The method happened
to be very efficient in application to different few-dimensional quantum problems
\cite{melezhik1999,melezhik2000,capel2003,melezhik2004}. Mathematical aspects of the 2D NDDVR were discussed
recently in \cite{carrington}.

In the 2D NDDVR we seek for the time-dependent solution $\psi(\rho_{R},{\bf r},t)$ of the problem (1-5)
according to the expansion \cite{melezhik1997,melezhik1999,melezhik2007}
\begin{equation}
\psi\!(\rho_{R},r,\Omega,t)= \sum_{j=1}^{N}f_j(\Omega)\,\psi_{j}(\rho_{R},r,t)\,
\end{equation}
with respect to the two-dimensional basis
\begin{equation}
f_j(\Omega)=\sum_{\nu=1}^{N}Y_{\nu}(\Omega)(Y^{-1})_{\nu j}\,\,
\end{equation}
associated with a mesh $\Omega_j = (\theta_{j_{\theta}},\phi_{j_{\phi}})$.  For the $\theta$
variable, the $N_{\theta}$ mesh points $\theta_{j_{\theta}}$
correspond to the zeros of the Legendre polynomial $P_{N_{\theta}}
(\cos \theta_{j_{\theta}})=0$. For the $\phi$ variable, the
$N_{\phi}$ mesh points are chosen as $\phi_{j_{\phi}} = \pi
(2j_{\phi}-1)/N_{\phi}$. The total number $N=N_{\theta}\times N_{\phi}$ of grid points
$\Omega_j = (\theta_{j_{\theta}},\phi_{j_{\phi}})$ is equal to the
number of basis functions in the expansion (6) and the number of terms in
the definition (7), where the symbol $\nu$ represents the
twofold index $\{l,m\}$ and the sum over $\nu$ is
equivalent to the double sum
\begin{equation} \sum_{\nu=1}^{N} =
\sum_{m=-(N_{\phi}-1)/2}^{(N_{\phi}-1)/2}\,\sum_{l=\mid
m\mid}^{\mid m\mid+N_{\theta}-1}\,.
\end{equation}
The $l$ and $m$ indexes show the number of zeros over $\theta$ and
$\phi$ variables of the polynomials $Y_{\nu}(\Omega)$ which we specify in the next paragraph.
Due to the definition (8) the values $N_{\phi}$ may be chosen only odd.
$N_{\theta}$ can take on arbitrary values. The coefficients
$(Y^{-1})_{\nu j}$ in the definition (7) are the elements of the $N\times N$ matrix
$Y^{-1}$ inverse to the matrix given by the values $Y_{j
\nu}=Y_{\nu}(\Omega_j)$ of the polynomials $Y_{\nu}(\Omega)$ at the grid points $\Omega_j$. It is clear
that $f_j(\Omega_{j'}) = \delta_{jj'}$ at such definition (i.e. the basis (7) belongs to the
class of Lagrange functions \cite{baye}) and the coefficients $\psi_{j}(\rho_{R},r,t)$ in the expansion
(6) are the values of the searching solution
$\psi(\rho_{R},{\bf r},t)$ at the points of angular grid $\Omega_j$:
$\psi_{j}(\rho_{R},r,t)=\psi(\rho_{R},r,\Omega_{j},t)$ .

The polynomials $Y_{\nu}(\Omega)$ in Eq.(7) are chosen as
\begin{equation} Y_{\nu}(\Omega) = Y_{lm}(\Omega) = e^{im\phi} \sum_{l'}
C^{l'}_{l}\times P^{m}_{l'}(\theta) \,,
\end{equation}
where
$C^{l'}_{l}=\delta_{ll'}$ holds in general, and thus $Y_{\nu}(\Omega)$
coincides with the usual spherical harmonic with a few possible
exceptions for large values of $\nu$ such
that we receive the orthogonality relation
\begin{equation} \int Y_{\nu}^{*} (\Omega) Y_{\nu'}
(\Omega) d\Omega \approx \sum_j \lambda_j Y_{\nu j}^{*} Y_{\nu' j} =
\delta_{\nu \nu'}
\end{equation}
for all $\nu$ and $\nu'\leq N$.
Here the $N$ weights $\lambda_j$ are the
standard Gauss-Legendre weights multiplied by $2\pi/N_{\phi}$.  For
most $\nu$ and $\nu'$ the above relation is automatically
satisfied because the basis functions $Y_{\nu} (\Omega)$ are
orthogonal and the Gaussian quadrature is exact. For these $\nu$
we have $ C^{l'}_{l} = \delta_{ll'}$ in Eq.(9).
However, in a few cases involving the highest $l$ values,
some polynomials $P^{m}_{l}(\theta)$ have to be orthogonalized
in the sense of the Gaussian quadrature ($C^{l'}_{l}\neq
\delta_{ll'}$ for these specific values of $l$). With this choice,
the matrix $S_{j\nu}=\lambda_j^{1/2} Y_{j \nu}$ is orthogonal.

The basis in the form (7) was initially introduced in \cite{melezhik1991} for the one-dimensional
angular space $\Omega=\theta$.
The fact, that values of the searching wave function $\psi(\rho_{R},{\bf r},t)$ at the
grid points of the 2D angular space $\psi(\rho_{R},r,\Omega_{j},t)=\psi_{j}(\rho_{R},r,t)$ are
utilized in the 2D NDDVR, drastically simplifies the calculations\cite{melezhik1997,melezhik1999,melezhik2007}
as compared to the usual partial-wave analysis.

\subsection{Splitting-up Method for 4D Time-Dependent Schr\"odinger Equation}
It is an attractive feature of the 2D NDDVR (6-10) that
for the grid representation $f_j(\Omega)$ the only nondiagonal part of the
Hamiltonian (1) is the angular part of the kinetic
energy operator (see Eqs.~(2,3))
$$
\{\frac{1}{2\mu
r^{2}}L^{2}(\theta,\phi)-\frac{1}{2M\rho_{R}^{2}}\frac{\partial^2}{\partial\phi^2}\}\sum_{\nu}^{N}
Y_{\nu}(\Omega)(Y^{-1})_{\nu
j'}\mid_{\Omega=\Omega_{j}}=
$$
$$
\sum_{\nu}^{N}Y_{j\nu}\{\frac{l(l+1)}{2\mu
r^{2}}+\frac{m^{2}}{2M\rho_{R}^{2}}\}(Y^{-1})_{\nu j'}\,
$$
which can be diagonalized by the simple unitary transformation $S_{j
\nu}=\lambda_j^{1/2} Y_{j \nu}$ \cite{melezhik1997,melezhik1998,melezhik1999}. This property has been
exploited for developing an efficient algorithm with a
computational time scaling proportional to the number $N=N_{\theta}\times
N_{\phi}$ of unknowns in the system of equations \cite{melezhik1998}
\begin{equation}
i\frac{\partial}{\partial t}\psi_j(\rho_R,r,t)=\sum_{j'}^N
H_{jj'}(\rho_R,r)\psi_{j'}(\rho_R,r,t)\,\,\,.
\end{equation}

For the propagation
$\psi_j(\rho_{R},r,t_{n})\rightarrow
\psi_j(\rho_{R},r,t_{n+1})$ in time $t_n\rightarrow
t_{n+1}=t_n+\Delta t$ we have developed a computational scheme \cite{melezhik2007} based on the component-by-component
split-operator method suggested by G.I.Marchuk in 1971 \cite{marchuk}. The Hamiltonian in (11) permits the
splitting into the following three parts
\begin{equation}
H_{jj'}(\rho_R,r) = h_{jj'}^{(0)}(\rho_{R}) +
h_{jj'}^{(1)}(r) + U_{j}(\rho_{R},\rho)\delta_{jj'} \,
\end{equation}
where
$$
h_{jj'}^{(0)}(\rho_{R}) = -\frac{\delta_{jj'}}{2M}(\frac{\partial
^2}{\partial \rho_{R}^{2}}+\frac{1}{4\rho_{R}^{2}})
+\frac{1}{2M\rho_{R}^{2}\sqrt{\lambda_{j}\lambda_{j'}}}\sum_{\nu}^{N}
(Y)^{-1}_{j\nu}m^{2}(Y^{-1})_{\nu j'} \,\,\, ,
$$
$$
h_{jj'}^{(1)}(r) = -\frac{\delta_{jj'}}{2\mu}\frac{\partial ^{2}}{\partial
r^{2}} +\delta_{jj'}V(r) - \frac{1}{2\mu
r^{2}\sqrt{\lambda_{j}\lambda_{j'}}}\sum_{\nu}^{N}
(Y)^{-1}_{j\nu}l(l+1)(Y^{-1})_{\nu j'} \,\,\, ,
$$
$$
U_{j}(\rho_{R},\rho)=
\frac{1}{2}(m_{1}\omega_{1}^{2}+m_{2}\omega_{2}^{2})\rho_{R}^{2}
+\frac{\mu^{2}}{2}(\frac{\omega_{1}^{2}}{m_{1}}+\frac{\omega_{2}^{2}}{m_{2}})\rho^{2}
+\mu(\omega_{1}^2-\omega_{2}^{2})\rho\rho_{R} cos \phi_{j} \,\,\,
.
$$
Subsequently we can approximate the time-step
$\psi_j(\rho_{R},r,t_{n})\rightarrow
\psi_j(\rho_{R},r,t_{n+1})$ where $t_n\rightarrow
t_{n+1}=t_n+\Delta t$ according to
\begin{equation} \psi(t_{n}+\Delta t) =
\exp(-\frac{i}{2}\Delta t \hat{U}) \exp(-i\Delta t\hat{h}^{(1)}) \exp(-i\Delta t
\hat{h}^{(0)})\times
\end{equation}
$$
\exp(-\frac{i}{2}\Delta t \hat{U})\psi(t_{n}) +
O(\Delta t^3 )\,\,\, .
$$

The time evolution proceeds as follows. For the first and the last
steps according to the relation (13) we write the function $\psi$
and the operators $\exp(-i\Delta t \hat{U}/2)$ in our 2D
NDDVR (7) on the 2D grid
$\{\Omega_{j}\}=\{\theta_{j_{\theta}},\phi_{j_{\phi}}\}$. Since the potential
$U_{j}(\rho_{R},\rho)$ is diagonal in this representation the
first and last steps represent simple
multiplications of the diagonal matrices $\exp(-\frac{i}{2}\Delta
t U_{j}(\rho_{R},\rho))$. Two intermediate steps in (13)
depending on $\hat{h}^{(0)}$ and $\hat{h}^{(1)}$ are treated in
the basis $Y_{\nu}$ (9) where the matrix operators
$\hat{h}^{(0)}(\rho_{R})$ and $\hat{h}^{(1)}(r)$ are diagonal with
respect to the indices $m$ and $l$. For that we approximate the
exponential operators according to
\begin{equation} \exp(-i\Delta t \hat{A}) \approx
(1+\frac{i}{2}\Delta t \hat{A})^{-1}(1-\frac{i}{2}\Delta t \hat{A}) +
O(\Delta t^{3})\, ,
\end{equation}
which ensures the desired accuracy of the
numerical algorithm (13). Thus, after the discretization of $r$ (or
$\rho_{R}$) with the help of finite-differences the matrix $\hat{A}$
possesses a band structure and we arrive at the following boundary-value
problems
$$
(1 + \frac{i}{2}\Delta t \hat{A})\psi(t_{n} + \frac{\Delta t}{4})
= (1 - \frac{i}{2}\Delta t \hat{A})\psi(t_{n})\,\,\, ,
$$
which can be solved rapidly due to the band structure of the
matrix $\hat{A}$. This computational
scheme is unconditionally stable \cite{marchuk}, preserves unitarity and
is very efficient, i.e. the computational time is proportional to the
total number $N$ of grid points over the radial and angular variables \cite{melezhik1998}.
The efficiency of the computational procedure is based on the fast transformation with
help of the unitary matrix
$S_{j \nu}=\lambda_j^{1/2} Y_{j \nu}$ between
the two relevant representations: the 2D NDDVR (7) and the $Y_{\nu}$-representation (9).

\subsection{2D Multichannel Scattering Problem as Boundary-Value Problem}

In the case of the pair collisions of two identical atoms in harmonic trap the
problem reduces to scattering
of a single effective particle with the reduced mass $\mu$, off a scatterer $V(r)$ at the origin,
under transverse harmonic confinement with frequency $\omega$ ($\hbar=1$)
\cite{saeidian2008}
\begin{equation}\label{poth}
\left [-\frac{1}{2\mu}\nabla_r^2 +
\frac{1}{2}\mu\omega^2\rho^2 + V(r)\right ] \psi (\mathbf{r}) =
\epsilon\psi(\mathbf{r})\,\,.
\end{equation}
Here the energy of the relative two-body motion $\epsilon =\epsilon^{(n,m)}_{\perp}+\epsilon_{\|}$ is a sum of the
transverse $\epsilon^{(n,m)}_{\perp}$ and longitudinal collision
$\epsilon_{\|}$ energies. Due to our definition of the confining
potential,  the transverse excitation energies $\epsilon^{(n,m)}_{\perp}$
can take the possible values $\epsilon^{(n,m)}_{\perp}= \epsilon-\epsilon_{\|}= \omega (2n + |m| +1) > 0$
of the discrete spectrum of the 2D oscillator $\frac{1}{2}\mu\omega^2\rho^2$.

The problem was in integration of the 3D stationary Schr\"odinger equation
(15) with the scattering asymptotic conditions
\begin{equation}\label{asymp}
\psi_{n,m}(\mathbf{r})=e^{ik_{n}z}\phi_{n,m}(\rho ,\varphi)+\sum_{n'=0}^{n_{o}}f_{nn'} e^{ik_{n'}|z|}\phi_{n',m}(\rho ,\varphi)
\end{equation}
at $\mid z\mid\rightarrow +\infty$
and finding the unknown matrix elements $f_{nn'}$ of the scattering
amplitude, which describe transitions between the initial $n$ and
final $n'$ channels of scattering. The quantum numbers $n$ of the two-dimensional harmonic oscillator denote
the transversal excitations/deexcitations $n\leftrightarrow n'$ of colliding
atoms. In addition to the quantum number $n$ the asymptotic scattering
state is defined also by the momentum $k_n$ of the channel
$$
k_n=\sqrt{2\mu\epsilon_{\|}}\,\,.
$$
The energy $\epsilon$ defines the number $n_o$ of open channels in (\ref{asymp}), where
the longitudinal collision energy must be positive $\epsilon_{\|}=\epsilon-\epsilon^{(n,m)}_{\perp} > 0$ .

It is clear that the scattering amplitude depends also on the index $m$ which, however, remains unchanged
during the collision due to the axial symmetry of the problem. Hereafter we consider only the case $m = 0$
and the index is omitted in the following. Due to the axial symmetry of
the interactions in (\ref{poth}) we separate the $\phi$-variable
and reduce the problem to 2D one in the variables $(r,\theta)$ (or $(z,\rho)$).

First, we discretize the 2D Schr\"{o}dinger equation
(\ref{poth}) on a 2D grid of
angular $\{\theta_j\}_{j=1}^{N_\theta}$ and radial $\{r_j\}_{j=1}^N$
variables. The angular grid points $\theta_j$ are defined as
the zeroes of the Legendre polynomial
$P_{N_{\theta}}(\cos\theta)$ of the order $N_{\theta}$. Using the completeness
property of the normalized Legendre polynomials which remains valid also on the chosen angular grid
\begin{equation}\label{completeness}
\sum^{N_{\theta} -1}_{l=0}P_l(\cos \theta_j)P_l(\cos
\theta_{j'})\sqrt{\lambda_j\lambda_{j'}}=\delta_{jj'}\,\,,
\end{equation}
where $\lambda_j$ are the weights of the Gauss quadrature, we
expand the solution of equation (\ref{poth}) in the basis
$f_j(\theta)=\sum^{N_{\theta} -1}_{l=0}P_l(\cos \theta)(\mathbf{P}^{-1})_{lj}$ according to
\begin{equation}\label{grid expansion of wave function}
\psi(r, \theta) = \frac{1}{r}\sum ^{N_{\theta
}}_{j=1} f_j(\theta) u_j(r)\,\,.
\end{equation}
Here $\mathbf{P}^{-1}$ is the inverse of the $N_{\theta}\times
N_{\theta}$ matrix $\mathbf{P}$ with the matrix elements defined
as $\mathbf{P}_{jl}=\sqrt{\lambda_j}P_l(\cos \theta_j)$. Due to
this definition one can use the completeness relation
(\ref{completeness}) in order to determine the matrix elements
$(\mathbf{P^{-1}})_{lj}$ as
$(\mathbf{P^{-1}})_{lj}=\sqrt{\lambda_j}P_l(\cos
\theta_j)$. It is clear from (\ref{grid expansion of wave
function}) that the unknown coefficients $u_j(r)$ in the
expansion are the values $\psi(r,\theta_j)$ of the
two-dimensional wave function $\psi(r,\theta)$ at the grid points
$\theta_j$ multiplied by $\sqrt{\lambda_j} r$. Near the origin $r\rightarrow 0$ we have $u_j(r)\simeq
r \rightarrow 0$ due to the definition (\ref{grid expansion of
wave function}) and the demand for the probability distribution
$\mid\psi(r,\theta_j)\mid^2$ to be bounded. Substituting
(\ref{grid expansion of wave function}) into (\ref{poth}) results a system of $N_{\theta}$
Schr\"{o}dinger-like coupled equations with respect to
the $N_{\theta}$-dimensional unknown vector
$\mathbf{u}(r)=\{\lambda^{1/2}_ju_j(r)\}^{N_{\theta}}_1$
\begin{equation}\label{Sheroedinger-like equation}
[\mathbf{H}^{(0)}(r)+2(\epsilon\mathbf{I}-\mathbf{V}(r))]\mathbf{u}(r)=0,
\end{equation}
where
\begin{equation}
\mathbf{H}^{(0)}_{jj'}(r)=\frac{d^2}{dr^2}\delta_{jj'}-\frac{1}{r^2}\sum^{N_{\theta}-1}_{l=0}
\mathbf{P}_{jl}l(l+1)(\mathbf{P}^{-1})_{lj'}\,\,, \end{equation}
\begin{equation}
\mathbf{V}_{jj'}(r)=V(r,\theta_j)\delta_{jj'}=\{V(r)+
\frac{1}{2}\omega^2\rho^2_j\}\delta_{jj'},\quad \rho_j=r\sin \theta_j\,\,,
\end{equation}
and $\mathbf{I}$ is the unit matrix.
We solve the system of equations (\ref{Sheroedinger-like equation}) on the quasi-uniform radial
grid \cite{melezhik1997}
\begin{equation}\label{r-mapping}
r_j = R\frac{e^{\gamma x_j}-1}{e^{\gamma}-1}\,\,,\,\,j=1,2,...,N
\end{equation}
of $N$ grid points $\{r_j \}$ defined by mapping $r_j\in [0,R\rightarrow
+\infty]$ onto the uniform grid $x_j\in [0,1]$ with the equidistant
distribution $x_j - x_{j-1} = 1/N$. By varying
$N$ and the parameter $\gamma > 0$ one can choose more adequate
distributions of the grid points for specific interatomic and
confining potentials.

By mapping the initial variable $r$ in Eq.(\ref{Sheroedinger-like equation}) onto $x$ we obtain
\begin{equation}\label{Sheroedinger-like equation2}
[\mathbb{H}^{(0)}(x)+2\{\epsilon\mathbf{I}-\mathbf{V}(r(x))\}]\mathbf{u}(r(x))=0\,\,,
\end{equation}
with
\begin{equation}\label{H^02}
\mathbb{H}^{(0)}_{jj'}(x)=f^2(x)\delta_{jj'}\left (\frac{d^2}{dx^2}-\gamma\frac{d}{dx} \right )
-\frac{1}{r^2(x)}\sum^{N_{\theta}-1}_{l=0}\mathbf{P}_{jl}l(l+1)(\mathbf{P}^{-1})_{lj'}\,\,, \end{equation}
where
\begin{equation}
f(x) = \frac{e^{\gamma} - 1}{Re^{\gamma x}\gamma}\,\,.
\end{equation}
The uniform grid with respect to $x$ gives  6-order accuracy for applying a
7-point finite-difference approximation of the derivatives in the
equation (\ref{Sheroedinger-like equation2}) . Thus, after the
finite-difference approximation the initial 2D Schr\"{o}dinger
equation (\ref{poth}) is reduced to
the system of N algebraic matrix equations
$$
\sum_{p=1}^3\mathbb{A}^j_{j-p}\mathbf{u}_{j-p}+
[\mathbb{A}^j_j+2\{\epsilon\mathbf{I}-\mathbf{V}_j\}]\mathbf{u}_j+\sum_{p=1}^3\mathbb{A}^j_{j+p}\mathbf{u}_{j+p}=0
\,\,,\,\, j=1,2,...,N-3
$$
\begin{equation}\label{LU}
\mathbf{u}_j+\alpha_j^{(1)}\mathbf{u}_{j-1}+\alpha_j^{(2)}\mathbf{u}_{j-2}+
\alpha_j^{(3)}\mathbf{u}_{j-3}+\alpha_j^{(4)}\mathbf{u}_{j-4}=\mathbf{g}_j\quad\quad j=N-2, N-1, N
\end{equation}
where each coefficient $\mathbb{A}^j_{j'}$ is a
$N_{\theta}\times N_{\theta}$ matrix, each $\alpha_j$ is a diagonal $N_{\theta}\times N_{\theta}$ matrix and each
$\mathbf{g}_j$ is a $N_\theta$-dimensional vector. Here the functions
$\mathbf{u}_{-3}$, $\mathbf{u}_{-2}$, $\mathbf{u}_{-1}$ and
$\mathbf{u}_{0}$ in the first three equations of the system (for
$j=1,2$ and 3) are eliminated by using the ``left-side'' boundary
conditions: $\mathbf{u}_{0}=0$ and $\mathbf{u}_{-j}=\mathbf{u}_{j}$
($j=1,2,3$). The last three equations in this system for $j=N,N-1$ and $N-2$ are the ``right-side'' boundary
conditions approximating at the edge points $r_{N-2},r_{N-1}$ and $r_N=R$ of the radial grid, the
scattering asymptotics (\ref{asymp}) for the desired wave function
$\mathbf{u}(r_j)$. In order to construct the ``right-side'' boundary
conditions (\ref{LU}) at $j=N-2,N-1$ and $N$ we used an idea
of ref.\cite{hu2003} i.e. the asymptotic behaviour
(\ref{asymp}) at the edge points $r_{N-2},r_{N-1}$ and $r_N=R$ are considered as
a system of vector equations with respect to the unknown vector
$f_{nn'}$ of the scattering amplitude for a fixed $n$. By
eliminating the unknowns $f_{nn'}$ from this system we implement the
``right-side'' boundary conditions defined by Eqs.(\ref{LU}) at $j=N-2,N-1$ and
$N$.

The reduction of the 2D multi-channel scattering problem to the
finite-difference boundary value problem (\ref{LU}) permits one to
apply efficient computational methods. Here we use
the $LU$-decomposition \cite{press} or sweep method \cite{gelfand}, which is a fast implicit
matrix algorithm. The
block-diagonal structure of the matrix of the coefficients in the system
of equations(\ref{LU}) with the width of the diagonal band equal
to $7\times N_\theta$ makes this computational scheme an efficient one.

Solving the problem (\ref{LU}) for the defined initial vector $k_n$ and
a fixed $n$ from the possible set $0\leq n\leq n_{e}$  we first calculate the vector function
$\psi(k_n,r,\theta_j)$. Then, by  matching the calculated vector $\psi(k_n,R,\theta_j)$ with the asymptotic behaviour
(\ref{asymp}) at $r=R$, we calculate the $n$-th row of the scattering amplitude matrix
$f_{nn'}$ describing all possible transitions $n\rightarrow
n'=0,1,...,n_{e}$. This procedure is repeated for the next
$n$ from $0\leq n\leq n_{o}$ and over upon calculating all the elements $f_{nn'}$ of the
scattering amplitude.

\section{Some Results}

In this section we briefly discuss some of the  most interesting results obtained
with the developed computational methods.

\subsection{Resonant molecule formation in harmonic waveguides}

In our work \cite{melezhik2009} it was shown that the quantum dynamics of the coupled
center-of-mass and relative motion for two different colliding atoms in a harmonic trap
exhibits a resonant formation process of ultracold molecules. With the developed splitting-up method (section 2.3),
permitting to treat quantitatively the collision in the 4D configuration space
$(\rho_R,r,\theta,\phi)$, we have analyzed in detail the quantum dynamics
of the process and demonstrated the existence of a novel mechanism for
the resonant formation of polar molecules. The origin of this effect is the confinement-induced
mixing (4) of the relative and center-of-mass motions in the atomic collision process
leading to a coupling of the diatomic continuum to center-of-mass excited molecular states in
closed transverse channels. The process is illustrated by Fig.1 where the probability density  distribution averaged over
the angular variables $W(\rho_R,r,t) = \int|\psi(\rho_R,r,\theta,\phi,t)|^2 (r^2\rho_R)^{-1} \sin\theta d\theta d\phi$ is given
as a function of time.
\begin{figure}
\centering
\includegraphics[width=17.cm]{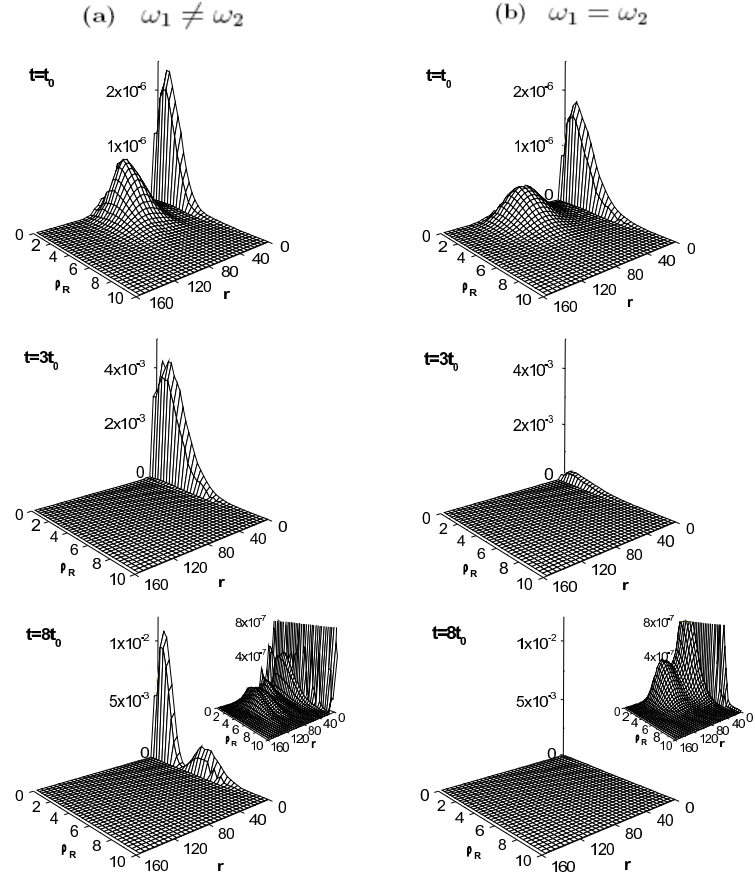}
\caption{Time evolution of the probability density  distribution averaged over
the angular variables $W(\rho_R,r,t) = \int|\psi(\rho_R,r,\theta,\phi,t)|^2 (r^2\rho_R)^{-1} \sin\theta d\theta d\phi$.
({\bf a}) For the waveguide with $\omega_1/2.2=\omega_2=0.02$
({\bf b}) for $\omega_1=\omega_2=0.02$. For $t=8t_0$ corresponding to a time
after the collision the insets show a more detailed view of $W$ on the scales
$0\leq W \leq 8\times 10^{-7}$ corresponding to the continuum part of the wave-packet.
$\epsilon_{\|}=0.004$, time is given in units of $t_0=\pi/\omega_2$.}
\label{fig1}
\end{figure}
For $\omega_1 \neq \omega_2$ (different atoms ``feel" different frequencies $\omega_1$ and $\omega_2$ in the trap)
we observe that a considerable part of the scattered wave-packet (see the probability density distribution
$W(\rho_R,r,t)$ at $t=8t_0$) is located near origin $r=0$ after the collision and
corresponds to a molecular bound state with excitation of the molecular center-of-mass. It is first excited state with
respect to the center-of-mass motion what is indicated by the node of the density distribution $W$ at the point about
$\rho_R\backsimeq 3$. Note, that during collision
at  $t\sim 3t_0$ the main part of the wave-packet is temporarily in the
ground state, decays thereafter rapidly into the
continuum but part of it goes into the excited molecular state.
In contrast to this the case $\omega_1 = \omega_2$
in Fig.1b (the case of the center-of-mass separation) shows an almost complete decay the molecular ground state
into the continuum: The remaining minor part of the probability density distribution near $r=0$ representing the
molecular part is much smaller compared to the case
$\omega_1 \ne \omega_2$.

It was also shown that the molecular formation probabilities can be
tuned by changing the trap frequencies $\omega_1$ and $\omega_2$ characterizing the transverse
modes of the atomic species in the trap.

\subsection{Multichannel Scattering and Confinement-Induced Resonances}
\begin{figure*}
\centering
\includegraphics[height=6cm,width=6cm]{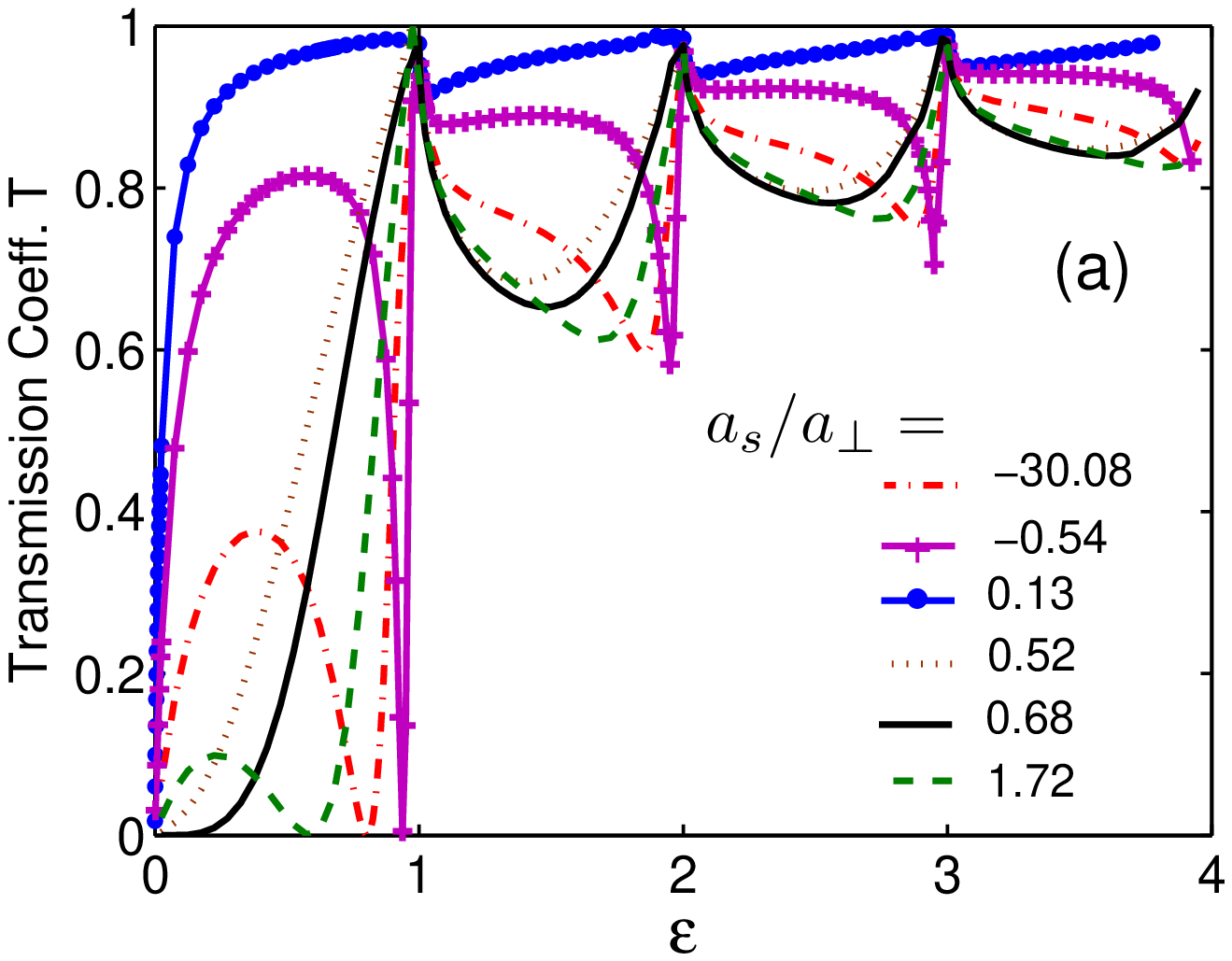}\includegraphics[height=6cm,width=6cm]{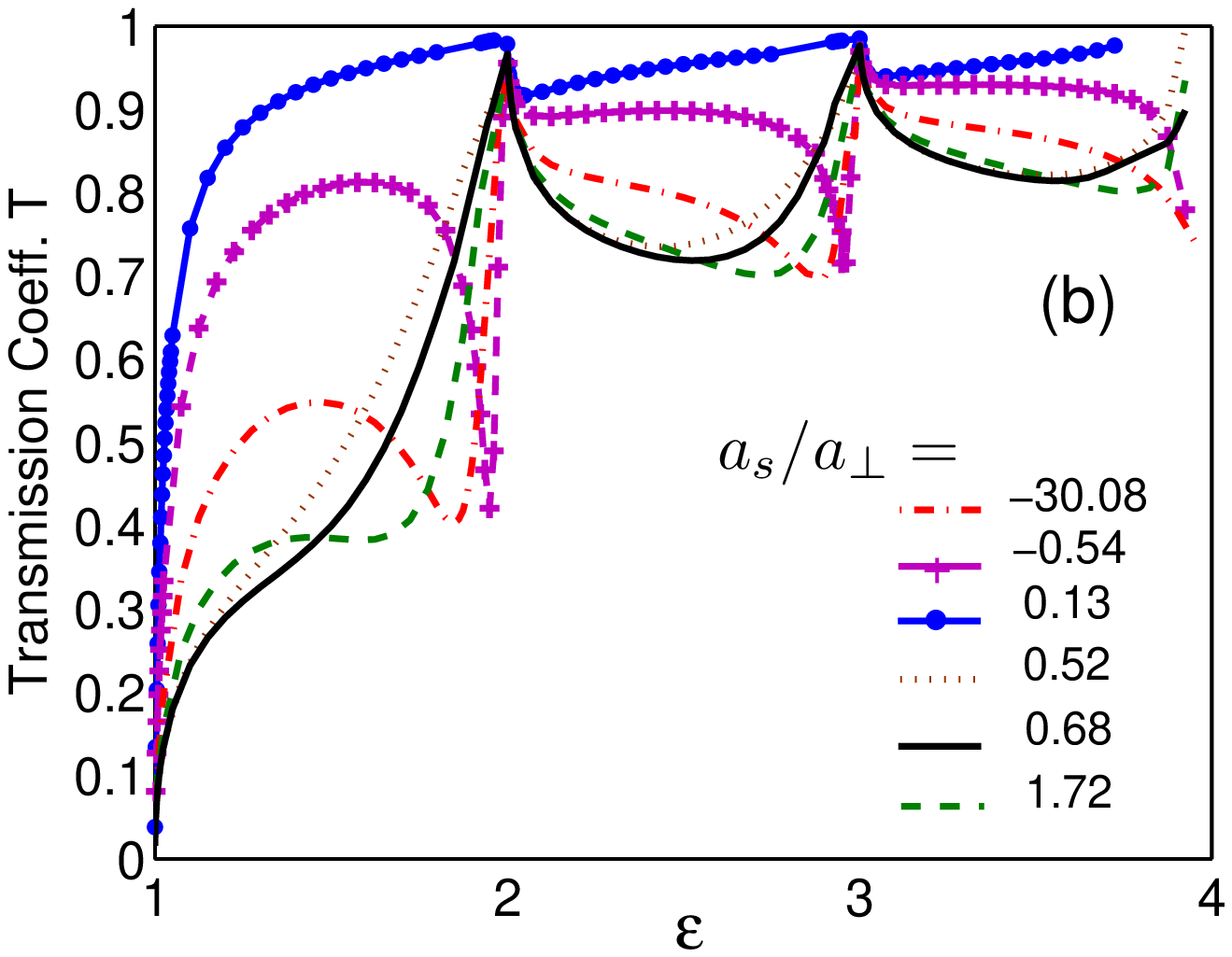}\\
\caption{ The total transmission coefficients $T$ for bosonic collisions as a function of the
dimensionless energy $\varepsilon=(\epsilon-\omega)/\omega$ for the two cases of the system being initially in (a) the ground transverse state $n=0$
and (b) the first excited transverse state $n=1$, for several ratios of $a_s/a_{\bot}$ and
$\omega = 0.002$. The black solid curve corresponds to $a_s/a_{\bot}=0.68$ for which
the zero-energy CIR in the single-mode regime is encountered \cite{olshanii}.} \label{fig6}
\end{figure*}
\begin{figure*}
\centering
\includegraphics[height=6cm,width=6cm]{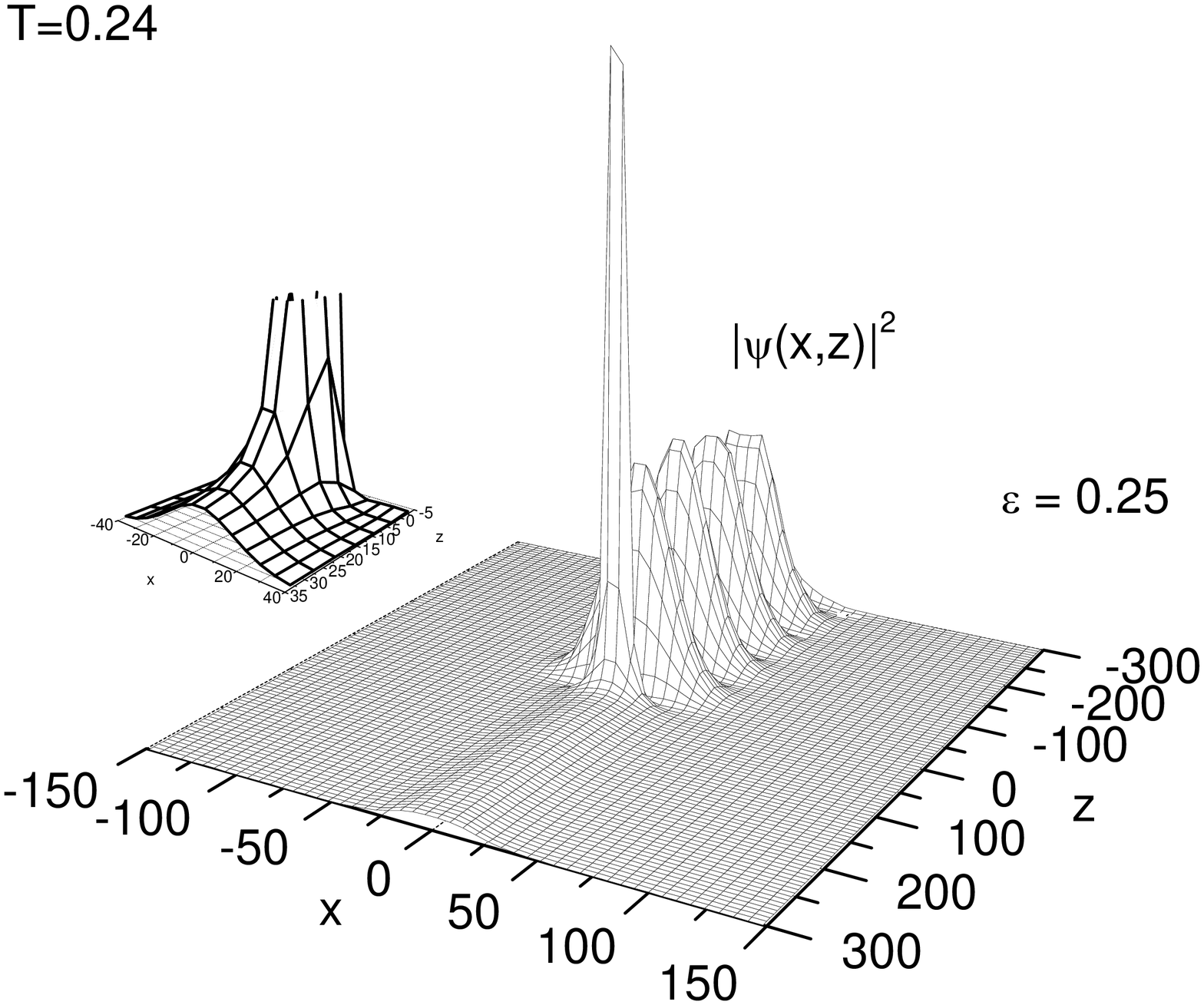}\includegraphics[height=6cm,width=6cm]{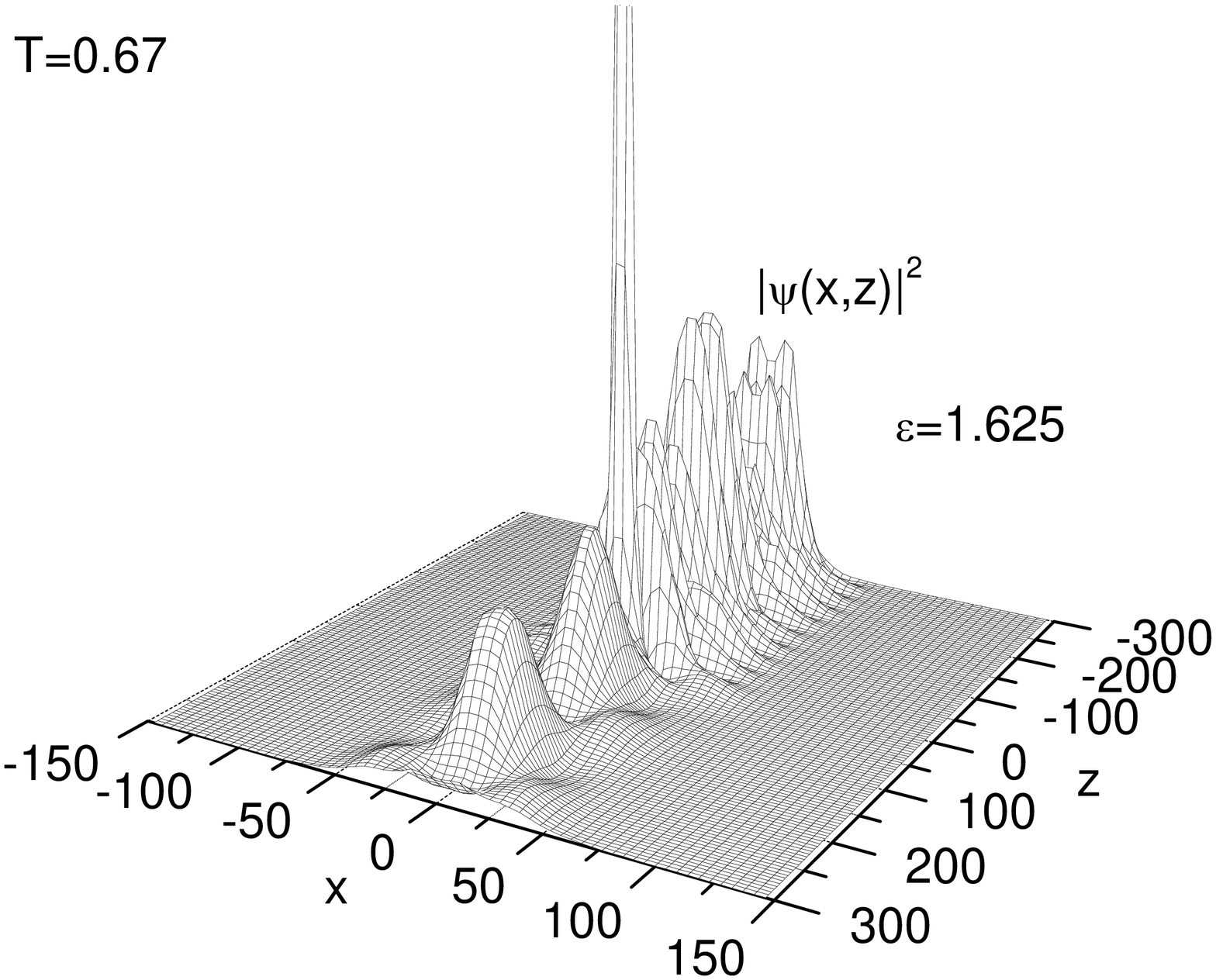}\\
\includegraphics[height=6cm,width=6cm]{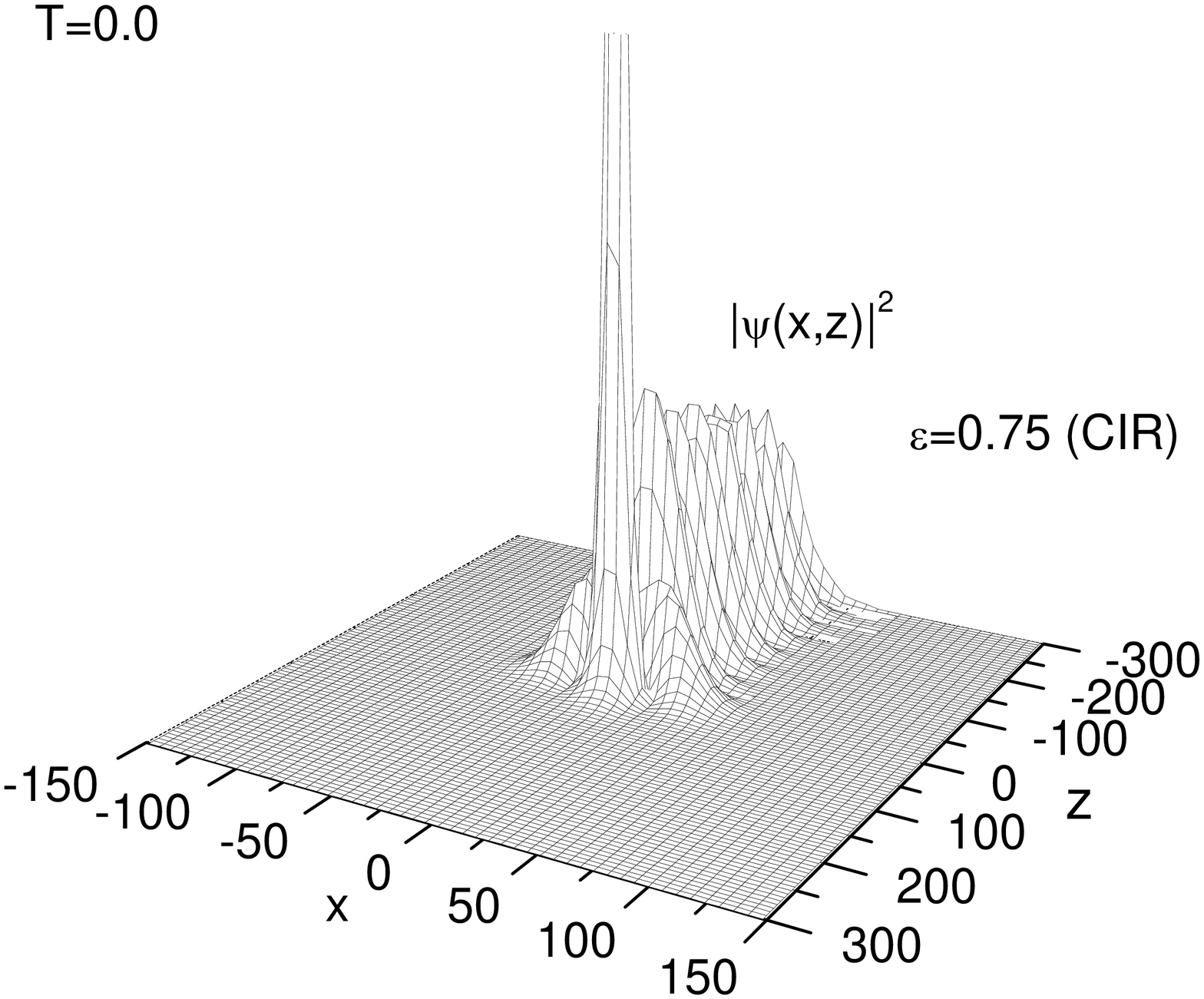}\includegraphics[height=6cm,width=6cm]{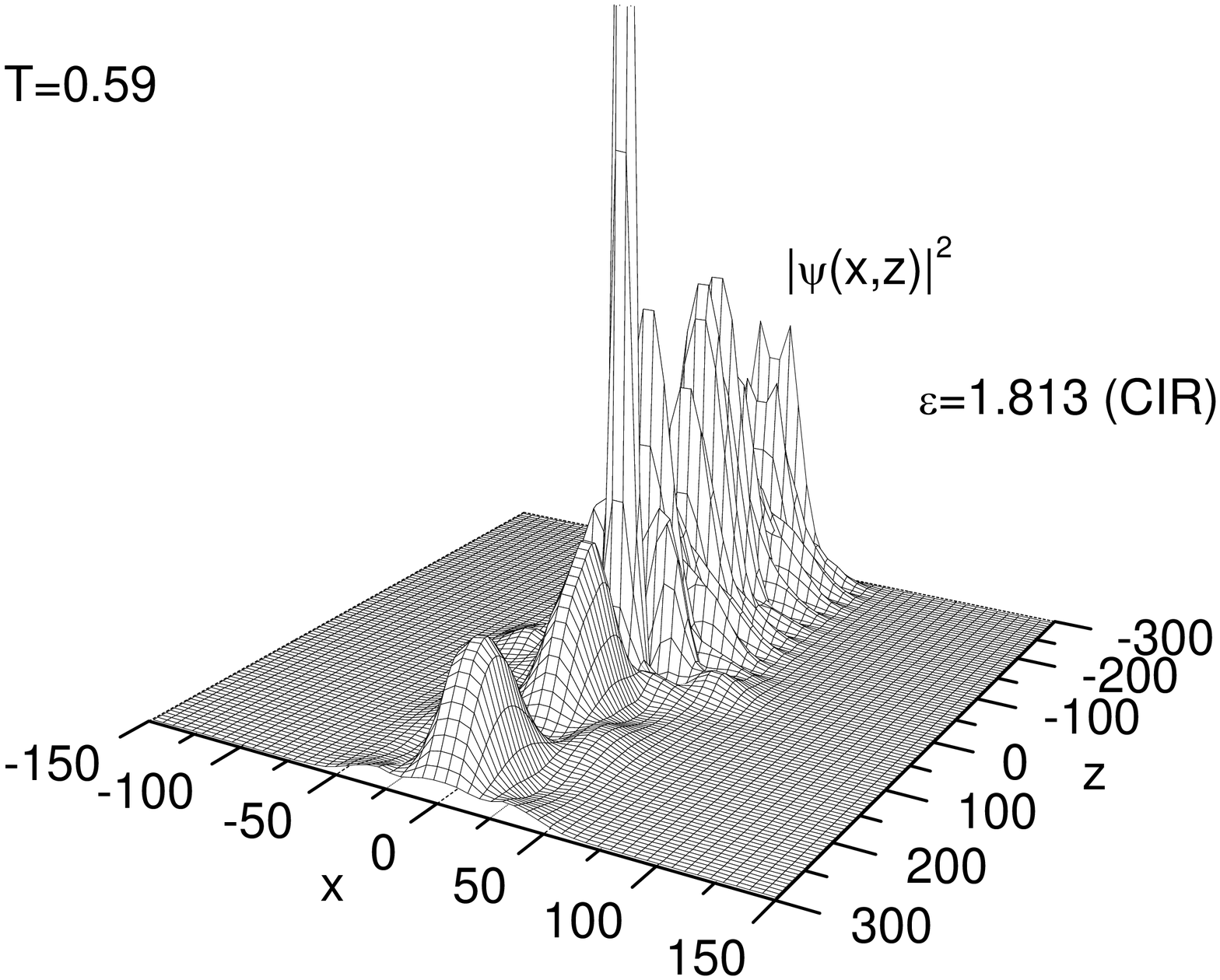}\\
\caption{The probability density $|\psi(x,z)|^2$ for bosonic collisions as a function of $x$ and $z$
at $a_s/a_{\bot}=+4.39$ for two cases of
the single-mode regime (left subfigures with $\varepsilon <1$) and two-mode regime (right subfigures with $1 <\varepsilon <2$).
The corresponding transmission values $T$ are also indicated. All subfigures are for $\omega=0.002$ and $n=0$.
The inset shows a more detailed view of $|\psi(x,z)|^2$ near the origin $z=x=0$.} \label{figE6}
\end{figure*}
The grid method for integration of the 2D multichannel scattering problem as a boundary-value problem (section 2.4) was applied to
extensively analyze the transverse excitations
and deexcitations as well as resonant scattering processes at pair atomic collisions in harmonic waveguiedes \cite{saeidian2008}.
Collisions of identical bosonic and fermionic as well as distinguishable
atoms in harmonic traps with a single frequency $\omega$
permitting the center-of-mass  separation were explored in depth.
In the zero-energy limit and single mode regime the well-known CIRs for bosonic \cite{olshanii,bergeman},
fermionic \cite{blume} and heteronuclear \cite{kim2006,kim2007} collisions  were reproduced.
In case of the multi-mode regime up to four open transverse channels were considered.
Important scattering observable in the quasi-1D scattering is the transmission coefficient
$T_n(\epsilon)=\sum_{n'}\frac{k_{n'}}{k_n}\mid\delta_{nn'}+f_{nn'}\mid^2$.

Fig.2 illustrates the behaviour of the
transmission $T$ with varying energy $\epsilon$ (permitting opening of the four lowest channels during the collision)
and scattering length $a_s/a_{\perp}$ (here $a_s$ is the scattering length in the 3D free space and $a_{\perp}=\sqrt{1/(\mu\omega)}$ is the
width of the trap). In Fig.3 the probability density distribution $\mid\psi(x,z)\mid^2$ calculated as a function of the the
transverse $(x)$ and longitudinal $(z)$ variables is given. Here the case $T=0$ illustrates the appearing of the CIR in the scattering.
The dual CIR \cite{kim2006,kim2007} leading to a complete quantum suppression of atomic scattering
was also analyzed in multi-channel regimes and possible applications were discussed.

\section{Conclusion}

In this lecture the computational methods have been considered which were elaborated for
a quantitative analysis of multichannel effects in low-dimensional few-body
systems. The efficiency of the methods was demonstrated in applications for
different quantum dynamics of two-body atomic collisions in harmonic 1D traps.
It is very promising further development of these approaches in
applications to hot problems of the few-body physics in confined geometry of optical
traps. Thus, the description of the effects of anisotropy and anharmonicity of the traps as well as
strong anisotropy in the interparticle interactions (scattering of polar
molecules) requires an extension to the increasing number of spatial variables.
Progress in this direction also opens up new possibilities for a quantitative analysis
of more complex low-dimensional few-body systems.

\end{document}